\documentstyle[12pt]{article}
\begin{document}


\centerline{\huge Convergence of the Schwinger --- DeWitt expansion}
\centerline{\huge for some potentials}
\vspace{1cm}
\centerline{{\Large V.~A.~Slobodenyuk}
			     \footnote{Physical--Technical Department,
			     Ulyanovsk State University,
			     432700 Ulyanovsk, Russian Federation.\\
			     E-mail: slob@themp.univ.simbirsk.su} }

\vspace{3cm}
{\large {\bf Running head:} Convergence of the Schwinger --- DeWitt
 expansion \dots}

\vspace{4cm}
{\bf Address of author:} \\
{\it Physical--Technical Department} \\
{\it Ulyanovsk State University } \\
{\it L.Tostogo str. 42, 432700 Ulyanovsk} \\
{\it Russian Federation} \\
{\it Tel.: (8422) 32-06-12} \\
{\it E-mail: slob@themp.univ.simbirsk.su}

\newpage

\begin{abstract}
It is studied time dependence of the evolution operator kernel
for the Schr\"odinger equation with a help of the Schwinger ---
DeWitt expansion. For many of potentials this expansion is
divergent.
But there were established nontrivial potentials for
which the Schwinger --- DeWitt expansion is convergent. These are,
e.g.,
 $$V=g/x^2, \quad V=-g/\cosh^2 x, \quad V=g/\sinh^2 x, \quad
						   V=g/\sin^2 x.$$
For all of them the expansion is convergent when
$g=\lambda (\lambda -1)/2$ and $\lambda$ is integer. The theories
with these potentials have no divergences and in this meaning
they are "good" potentials contrary to other ones.  So, it seems
natural to pay special attention namely to these "good" potentials.
Besides convergence they have other interesting feature: convergence
takes place only for discrete values of the charge $g$. Hence,
in the theories of this class the charge is quantized.
\end{abstract}

\medskip

\subsection*{Key words}

Asymptotic expansions, Schr\"odinger equation, evolution operator kernel,
quantization of charge.

\bigskip

\section{General relations}

Various approaches in the quantum theory use the short-time
Schwinger --- DeWitt expansion~[Schwinger 1951, DeWitt 1965, 1975,
Osborn 1983, Barvinsky 1995].
So as other expansions in different parameters: conventional
perturbation theory~[Bender 1969, Lipatov 1977], the WKB-expansion,
$1/n$-expansion~[Popov 1992]
etc, it is usually treated as asymptotic one.
Because of divergence of these expansions many difficulties in
the quantum theory arises. Particularly, in QCD one cannot
obtain correct predictions for low energy phenomena and so on.

Now we consider one interesting phenomenon, related to the
Schwinger --- DeWitt expansion, which, may be, will allow
us in the future to take of some problems arising from
divergence of the series.

The Schwinger --- DeWitt expansion is specific representation
of solution of the following problem for the evolution
operator kernel
  \begin{equation} \label{f1}
  i\frac{\partial}{\partial t} \langle q',t\mid q,0 \rangle =
  -\frac{1}{2} \frac{\partial^2}{\partial q'^2}
  \langle q',t\mid q,0 \rangle + V(q') \langle q',t\mid q,0 \rangle,
  \end{equation}
with initial condition
  \begin{equation} \label{f2}
  \langle q',t=0\mid q,0 \rangle = \delta (q'-q).
  \end{equation}
The kernel $\langle q',t\mid q,0 \rangle$ is written as
  \begin{equation} \label{f3}
  \langle q',t\mid q,0 \rangle = \frac{1}{\sqrt{2\pi it}}
  \exp \left\{i \frac{(q'-q)^2}{2t} \right\} F(t;q',q),
  \end{equation}
and $F$ according to~[Schwinger 1951, DeWitt 1965, 1975,
Slobodenyuk 1993, 1995~\cite{S2}] is expanded in powers of $t$
  \begin{equation} \label{f4}
  F(t;q',q) = \sum_{n=0}^{\infty} (it)^n a_n(q',q).
  \end{equation}
Here and everywhere below dimensionless values defined in obvious
manner are used. The potential $V(q)$ is continuous function.

It is easy to derive from~(\ref{f1})--(\ref{f3}) the problem
for the function $F$. The latter should satisfy the equation
  \begin{equation} \label{f5}
  i\frac{\partial F}{\partial t} =
  -\frac{1}{2} \frac{\partial ^2 F}{\partial q'^2} +
  \frac{q'-q}{it} \frac{\partial F}{\partial q'} + V(q')F
  \end{equation}
and initial condition
  \begin{equation} \label{f6}
  F(t=0;q',q)=1.
  \end{equation}

The coefficient functions $a_n(q',q)$ may be determined from
the sequence of recurrent relations
  \begin{equation} \label{f7}
  a_0(q',q)=1,
  \end{equation}
  \begin{equation} \label{f8}
  na_1(q',q) + (q'-q) \frac{\partial a_1(q',q)}{\partial q'} =
                                       a_1(q',q')= -V(q'),
  \end{equation}
for $n>1$
  \begin{equation} \label{f9}
  na_n(q',q) + (q'-q) \frac{\partial a_n(q',q)}{\partial q'}
  = \frac{1}{2} \frac{\partial ^2 a_{n-1}(q',q)}{\partial q'^2} -
  V(q')a_{n-1}(q',q).
  \end{equation}
Eqs.~(\ref{f7})--(\ref{f9}) show that $a_n$ can be calculated
via $a_{n-1}$ by means of integral relations
  \begin{equation} \label{f10}
  a_n(q',q)= \int \limits_0^1 \eta^{n-1} d\eta
  \left\{ \frac{1}{2} \frac{\partial ^2}{\partial x^2} -
  V(x) \right\} a_{n-1}(x,q)\Biggl.
  \Biggr|_{x=q+(q'-q)\eta}.
  \end{equation}
Combinations of Eqs.~(\ref{f10}) for different numbers
$n$ allows us to represent $a_n$ for given $n$ through the
potential $V$ and its derivatives
  \begin{eqnarray} \label{f11}
  a_n(q',q)&=& -\int\limits_0^1 \eta_n^{n-1} d\eta_n
  \int\limits_0^1 \eta_{n-1}^{n-2} d\eta_{n-1} \dots
  \int\limits_0^1 \eta_2 d\eta_2 \int\limits_0^1 d\eta_1
  \times \nonumber \\ &&
  \left\{ \frac{1}{2} \frac{\partial^2}{\partial x_n^2} -V(x_n)\right\}
  \left\{ \frac{1}{2} \frac{\partial^2}{\partial x_{n-1}^2}
				   -V(x_{n-1})\right\} \dots
  \nonumber \\ &&
  \left\{\frac{1}{2} \frac{\partial^2}{\partial x_2^2}  -V(x_2)\right\}
  V(x_1).
  \end{eqnarray}
Here $x_i=q+(x_{i+1}-q)\eta_i$, $x_{n+1}=q'$. Derivatives with
respect to different $x_i$ may be easily connected with each
other
  \begin{equation} \label{f12}
  \frac{\partial}{\partial x_i}=\eta_{i-1} \frac{\partial}{\partial x_{i-1}}
  =\eta_{i-1} \eta_{i-2} \frac{\partial}{\partial x_{i-2}}
  \end{equation}
etc.

Other useful representation for the kernel may be obtained if
in some domain for potential $V(q)$ the Taylor expansion takes
place
  \begin{equation} \label{f13}
  V(q') = \sum_{k=0}^{\infty} \Delta q^k \frac{V^{(k)}(q)}{k!}
  \end{equation}
(here $V^{(k)}(q)$ is $k$th derivative of $V(q)$, $\Delta q =q'-q$),
then the following representation of $F$ may be used in
calculations
  \begin{equation} \label{f14}
  F(t;q',q) = 1+
  \sum_{n=1}^{\infty} \sum_{k=0}^{\infty} (it)^n \Delta q^k b_{nk}(q).
  \end{equation}
For the coefficients $b_{nk}$ one has algebraic recurrent relations:
  \begin{equation} \label{f15}
  b_{1k}=- \frac{V^{(k)}(q)}{(k+1)!},
  \end{equation}
  \begin{equation} \label{f16}
  b_{nk}=\frac{1}{n+k} \left[\frac{(k+1)(k+2)}{2} b_{n-1, k+2} -
  \sum_{m=0}^k \frac{V^{(m)}(q)}{m!} b_{n-1, k-m} \right].
  \end{equation}

Both representations~(\ref{f11}) and (\ref{f14})--(\ref{f16})
are used for analysis of convergence of the Schwinger ---
DeWitt expansion (behaviour of $a_n$ for $n \to \infty$) in
general case and for specific potentials.

This formalism can be easily
modified for application to singular potentials with
singularity of type $1/q^2$ at $q=0$~[Slobodenyuk 1996~\cite{MPLA2}].
One should take instead of initial condition~(\ref{f2})
the following one
  \begin{equation} \label{f17}
  \langle q',t=0\mid q,0 \rangle = \delta (q'-q) + A \delta (q'+q)
  \end{equation}
which may provide fulfillment of boundary condition for
the wave function $\psi(q)$ at $q=0$ ($\psi(q)$ should vanish
at $q=0$) by appropriate choice of constant $A$. Constant
$A$ is determined by requirement that the kernel does not have
singularity at $q=0$ or $q'=0$ ($t \ne 0$). In correspondence
with~(\ref{f17}) the kernel is represented through two
functions $F^{(\pm)}$ as
  \begin{eqnarray} \label{f18}
  \langle q',t\mid q,0 \rangle &=& \frac{1}{\sqrt{2\pi it}}
  \exp \left\{i \frac{(q'-q)^2}{2t} \right\} F^{(-)}(t;q',q) +
  \nonumber \\ &&
  A \frac{1}{\sqrt{2\pi it}}
  \exp \left\{i \frac{(q'+q)^2}{2t} \right\} F^{(+)}(t;q',q),
  \end{eqnarray}
where $F^{(\pm)}$ can be expanded analogously to~(\ref{f4}).

Generalization on the three-dimensional case is obvious. So
we will not discuss it specially.

\section{Examples of convergent expansions}

It was shown in~[Slobodenyuk 1995~\cite{S2}] that
$|a_n(q',q)| \sim \Gamma(bn)$,
where $0<b \le 1$, if there is no any cancellations between
different contributions into $a_n$. For the most number
of the potentials such factorial growth really takes place.
Namely, for $V(q)$ represented by polynomial of order $L$ constant
$b$ is $b=(L-2)/(L+2)$, for other $V(q)$ constant $b$ is equal to 1.
So, the Schwinger --- DeWitt expansion factorially diverges
and the point $t=0$ is essential singular point of the
function $F$.  But there exist
some kinds of potentials, for which cancellations are so
essential that the series~(\ref{f4}) is convergent, and
$F$ is analytic at the point $t=0$.
This takes place only for some discrete values of charge.
The examples of such potentials were presented
in~[Slobodenyuk 1995~\cite{S3}, 1996~\cite{TMF2}]
  \begin{equation} \label{f19}
  V(q) = - \frac{g}{\cosh^2 q},
  \end{equation}
  \begin{equation} \label{f20}
  V(q) = \frac{g}{q^2},
  \end{equation}
  \begin{equation} \label{f21}
  V(q) = \frac{g}{\sinh^2 q},
  \end{equation}
  \begin{equation} \label{f22}
  V(q) = \frac{g}{\sin^2 q},
  \end{equation}
  \begin{equation} \label{f23}
  V(q) = aq^2 + \frac{g}{q^2}.
  \end{equation}
For all of them the expansion is convergent when
$g=\lambda (\lambda -1)/2$ and $\lambda$ is integer.

For illustration we consider the potential~(\ref{f20}).
This potential has singularity at $q=0$, so special formalism
should be applied here. But for the sake of brevity we
calculate only the function $F^{(-)}$, which is denoted
simply as $F$.

Expansion~(\ref{f13}) for the potential~(\ref{f20}) has
the finite convergence range $R(q)=q$, finiteness of which
is connected with singularity of $V(q)$ at the point $q=0$.
The derivatives $V^{(k)}$ may be easily calculated
  \begin{equation} \label{f24}
  V^{(k)}(q)= (-1)^k \frac{\lambda (\lambda -1)}{2}
  \frac{(k+1)!}{q^{k+2}}.
  \end{equation}
Substituting~(\ref{f24}) into~(\ref{f15}), (\ref{f16}) and
diminishing $n$ times the first index of $b_{nk}^{(-)}$ by means
of~(\ref{f16}) we get
  \begin{eqnarray} \label{f25}
  b_{nk}^{(-)}&=& \frac{(-1)^{n+k}}{q^{2n+k}} \frac{(k+n-1)!}{n!(n-1)!k!}
	\prod_{j=1}^n \left(\frac{\lambda (\lambda -1)}{2}
	- \frac{j(j-1)}{2} \right)
  \nonumber \\ &=&
	\frac{(-1)^{n+k}}{q^{2n+k}} \frac{(k+n-1)!}{n!(n-1)!k!}
	\frac{\Gamma(\lambda+n)}{2^n \Gamma(\lambda -n)}.
  \end{eqnarray}
It is obvious that if $\lambda$ is noninteger then
$|b_{nk}^{(-)}| \sim n!$ for $n \to \infty$.
So, for noninteger $\lambda$ expansion~(\ref{f14})
for potential~(\ref{f20}) is divergent. But if $\lambda$ is
integer ($\lambda >1$, because cases
$\lambda =0, \ \lambda =1$ are trivial) then one can easy
see from~(\ref{f25}) that only the coefficients $b_{nk}^{(-)}$ for
$n< \lambda$ are
different from zero, and in~(\ref{f14}) the series in powers of $t$ is
really the polynomial of finite degree $\lambda -1$.
Let us substitute~(\ref{f25}) into~(\ref{f14}) and
make summation over $k$. Then we get finally
  \begin{equation} \label{f26}
  F^{(-)}(t;q',q)= 1+ \sum_{n=1}^{\infty}
	     \left( \frac{-it}{2q'q} \right)^n
	     \frac{\Gamma(\lambda+n)}{n! \Gamma(\lambda-n)}.
  \end{equation}
The series in~(\ref{f26}) has the following feature: if $\lambda$
is noninteger, then coefficient in front of $t^n$ growths as $n!$
for $n \to \infty$ and the series is asymptotic; but if $\lambda$
is integer, then series contains only finite number of terms
(sum is made really till $n=\lambda -1$). So, in latter case
the expansion is convergent.

Singularities of $F^{(\pm)}$ at $q=0$ cancel each other and the
kernel $\langle q',t\mid q,0 \rangle$ has no such singularity.
Substitution of obtained expressions for $F^{(\pm)}$ into~(\ref{f18})
gives us asymptotic expansion of the function
  \begin{equation} \label{f27}
  \langle q',t\mid q,0 \rangle = e^{-i\frac{\pi}{2}(\lambda-1/2)}
  \frac{\sqrt{q'q}}{it}
  \exp \left\{i \frac{q'^2+q^2}{2t}  \right\}
  J_{\lambda-1/2}\left( \frac{q'q}{t}\right)
  \end{equation}
for small $t$ (large $q'q/t$). Eq.~(\ref{f27}) coincides with
known expression derived by other methods. Asymptotic
expansion for the Bessel function is not divergent only for
semi-integer order, i.e., for integer $\lambda$. In this case
the series contains finite number of terms and point $t=0$ is
regular point contrary to the case of noninteger $\lambda$, when
$t=0$ is essential singular point.

For other potentials of the series~(\ref{f19})--(\ref{f23}) the
situation is similar. The expansion~(\ref{f4}) is convergent
only for integer $\lambda$ and divergent in other case (but it
includes infinite number of terms).

Moreover, careful study shows us that if we
consider the coupling constant $g$ of continuous potential
$V(q)$ as independent variable, then the coefficients $a_n$
of representation~(\ref{f3})--(\ref{f4})
for the evolution operator kernel increase for $n \to \infty$ as
  $$a_n \sim \Gamma \left( n \frac{L-2}{L+2} \right)$$
for the potentials being expressed via the polynomial of order
$L$ and as
  $$a_n \sim n!$$
for other potentials. So, the Schwinger --- DeWitt expansion
in this supposition
is divergent for all potentials excluding polynomials
of order not higher then two.

If the charge is treated as fixed parameter, then because of
cancellations for some kinds of the potentials and for some
values of the charge $g$ expansion~(\ref{f4}) may be
convergent. Examples of such potentials are presented
above. Discreteness of the charge for the class
of the potentials for which the Schwinger --- DeWitt
expansion is convergent, probably, may be connected with
discreteness of the charge in the nature. In this correspondence,
the potentials of this class are of special interest.
Operating with them we get rid of some kind of divergences
in the theory and, at the same time, have a theory with
discrete charge. So, it is necessary to look for other
potentials of this class and study them carefully.
Besides, it is interesting to propagate such analysis on the
quantum field theory. One may hope that it will allow us to
reconstruct quantum electrodynamics with exactly fixed charge $e$.

\section*{Acknowledgments}

It is pleasure to acknowledge the Deutsche Forschungsgemeinschaft
for support of my participation in the conference
"Quantum Structures' 96, Berlin" and K.-E.~Hellwig and M.~Trucks for
hospitality.

\end{document}